\documentclass{ws-mpla}

\begin{document}

\markboth{W.T.H. van Oers}
{FROM HADRONIC PARITY VIOLATION TO PARITY-VIOLATING ELECTRON...} 

\catchline{}{}{}{}{}

\title{FROM HADRONIC PARITY VIOLATION TO PARITY-VIOLATING ELECTRON SCATTERING AND TESTS OF THE STANDARD MODEL
\footnote{Work supported in part by NSERC (Canada) and TRIUMF} } 

\author{\footnotesize WILLEM T. H. VAN OERS}
 

\address{Department of Physics and Astronomy, University of Manitoba, \\ Winnipeg,
 MB, Canada R3T 2N2\\ 
and \\TRIUMF, 4004 Wesbrook Mall, Vancouver, BC, Canada V6T 2A3\footnote{email: vanoers@triumf.ca}}

\maketitle

\pub{Received (Day Month Year)}{Revised (Day Month Year)}

\begin{abstract}
    Searches for parity violation in hadronic systems started soon after the
   evidence for parity violation in $\beta$ -decay of $^{60}$Co was presented
   by Madame Chien-Shiung Wu and in $\pi$ and $\mu$ decay by Leon Lederman in
   1957. The early searches for parity violation in hadronic systems did not
   reach the sensitivity required and only after technological advances in
   later years was parity violation unambiguously established. Within the
   meson-exchange description of the strong interaction, theory and
   experiment meet in a set of seven weak meson-nucleon coupling
   constants. Even today, after almost five decades, the determination of the
   seven weak meson-nucleon couplings is incomplete. Parity violation
   in nuclear systems is rather complex due to the intricacies of QCD. More
   straight forward in terms of interpretation are measurements of the
   proton-proton parity-violating analyzing power (normalized differences in
   scattering yields for positive and negative helicity incident beams), for
   which there exist three precision experiments (at 13.6, at 45, and 221 MeV).
   To-date, there are better possibilities for theoretical interpretation using
   effective field theory approaches. 
   
   The situation with regard to the measurement of the parity-violating
   analyzing power or asymmetry in polarized electron scattering is quite
   different. Although the original measurements were intended to determine
   the electro-weak mixing angle, with the current knowledge of the
   electro-weak interaction and the great precision with which electro-weak
   radiative corrections can be calculated, the emphasis has been to study
   the structure of the nucleon, and in particular the strangeness content
   of the nucleon. A whole series of experiments (the SAMPLE experiment at
   MIT-Bates, the G0 experiment and HAPPEX experiments at Jefferson
   Laboratory (JLab), and the PVA4 experiment at MAMI) have indicated that the
   strange quark contributions to the charge and magnetization distributions
   of the nucleon are tiny. These measurements if extrapolated to zero
   degrees and zero momentum transfer have also provided a factor five
   improvement in the knowledge of the neutral weak couplings to the quarks.\\ 
   Choosing appropriate kinematics in parity-violating electron-proton
   scattering permits nucleon structure effects on the measured analyzing
   power to be precisely controlled. Consequently, a precise measurement
   of the `running' of $\sin^{2}\theta_W$ or the electro-weak mixing angle
   has become within reach. The $Q^{p}_{weak}$ experiment at Jefferson
   Laboratory is to measure this quantity to a precision of about 4\%.
   This will either establish conformity with the Standard Model of quarks
   and leptons or point to New Physics as the Standard Model must be
   encompassed in a more general theory required, for instance, by a
   convergence of the three couplings (strong, electromagnetic, and weak)
   to a common value at the GUT scale.\\
   The upgrade of CEBAF at Jefferson Laboratory to 12~GeV, will allow
   a new measurement of $\sin^{2}\theta_W$ in parity-violating
   electron-electron scattering with an improved precision to the current
   better measurement (the SLAC E158 experiment) of the `running' of
   $\sin^{2}\theta_W$ away from the $Z^{0}$ pole. Preliminary design studies
   of such an experiment show that a precision comparable to the most
   precise individual measurements at the $Z^{0}$ pole (to about $\pm 0.00025$)
   can be reached. The result of this experiment will be rather
   complementary to the $Q^{p}_{weak}$ experiment in terms of sensitivity
   to New Physics. 
\end{abstract}

\ccode{PACS Nos.: 11.30.En; 12.60.-i; 1420.Dh}

\section{Hadronic Parity Violation}

    Manifestations of the weak interaction of quarks have been
    searched for since Madame Chien-Shiung Wu presented the evidence
    for parity violation in $\beta$-decay of $^{60}$Co and Leon Lederman
    for parity violation in $\pi$ and $\mu$ decay. For instance,
    Neil Tanner shortly thereafter, studied the 340 keV resonance in the
    $^{19}F(p,\alpha)$ reaction to the $^{16}$O ground state forbidden by
    angular momentum and parity conservation.[1] The hadronic weak
    interaction is studied by observing non-leptonic flavor changing
    decays of mesons and baryons and by measuring observables that
    conserve flavor but that violate reflection symmetry of the strong
    and electromagnetic interactions. One approach that has been pursued
    is the study of parity forbidden transitions between nuclear states in
    particular in the light nuclei. It was realized that two accidental
    aspects of nuclear structure in certain nuclei could amplify the
    expected effects of parity violation by several orders of magnitude
    beyond the nominal $O(10^{-7})$. The amplification arises from the
    near-degeneracy of opposite parity states that are mixed by the
    hadronic weak interaction and from the interference of an otherwise
    by parity conservation forbidden transition amplitude with a much
    stronger parity allowed transition amplitude. Although the weak
    quark-quark interaction has been well established its appearance in
    nuclear systems is clouded by effects of nuclear structure and by the
    dynamics of quantum chromodynamics (QCD) in the non-perturbative
    regime. Within the meson exchange description of the strong
    interaction, theory and experiment meet in a set of seven weak
    meson-nucleon coupling constants as defined in the seminal paper by
    Desplanques, Donoghue, and Holstein (DDH). [2] It is assumed that the
    parity violating N-N interaction is governed by the exchange of the
    pion and the two lightest vector mesons. The seven weak meson nucleon
    coupling constants are designated $h^{1}_{\pi}$, $h^{0,1,2}_{\rho}$,
    $h^{1'}_{\rho}$, and $h^{0,1}_{\omega}$, where the superscript
    indicates the isospin and the subscript the exchanged meson. Of these
    $h^{1'}_{\rho}$ was omitted from further consideration due to the
    difficulty calculating it. Desplanques, Donoghue, and Holstein
    provided `reasonable ranges' and `best values' for the weak meson
    nucleon couplings constants. These are given in Table 1 together with
    values for the latter from two similar theoretical
    approaches (Dubovic and Zenkin [3], and Feldman et al. [4]). 

 \begin{table}[t!]
      \tbl{Theoretical `reasonable ranges' (second column) and `best values'
    columns (three to five) for the parity violating meson-nucleon coupling
    constants from Desplanques, Donoghue, and Holstein, from Dubovic and 
    Zenkin, and from Feldman et al.. All values are given in units of
    $g_{\pi} = 3.8 \times 10^{-8} \approx G_{F}F^{2}_{\pi}/(2\sqrt{2})$.}
     {\begin{tabular}{@{}lcccr@{}} \toprule
   Coupling constant & DDH range & DDH best value & DZ & FCDH \\
  \colrule
   $h^{1}_{\pi}$    & $  0 \rightarrow -30$ & $ +12$ & $ +3$ & $ +7$\\
   $h^{0}_{\rho}$   & $ 30 \rightarrow -81$ & $ -30$ & $-22$ & $-10$\\
   $h^{1}_{\rho}$   & $ -1 \rightarrow   0$ & $-0.5$ & $ +1$ & $ -1$\\
   $h^{2}_{\rho}$   & $-20 \rightarrow -29$ & $ -25$ & $-18$ & $-18$\\
   $h^{0}_{\omega}$ & $ 15 \rightarrow -27$ & $  -5$ & $-10$ & $-13$\\
   $h^{1}_{\omega}$ & $ -5 \rightarrow  -2$ & $  -3$ & $ -6$ & $ -6$\\   \botrule     
  \end{tabular}}
   \end{table}

    The experimental results from the nuclear parity violating measurements
    have been analyzed using the framework of Desplanques, Donoghue, and
    Holstein leading to constraints on combinations of the weak meson nucleon
    coupling constants. There is a measure of agreement with the theoretical
    `reasonable ranges' although also experimentally the allowed ranges of
    values are large. Note that there exist two experimental inconsistencies
    as depicted in Fig. 1: the value of $h^{1}_{\pi}$ deduced from the
    $\gamma$-decays of $^{18}$F is consistent with zero while the result
    deduced from the anapole moment of $^{133}$Cs indicates a difference
    from zero with several standard deviations. The other inconsistency is
    possibly in the results deduced from the anapole moments of $^{205}$Tl
    and $^{133}$Cs. 
   \begin{figure}
\centerline{\includegraphics*[width=0.7\linewidth]{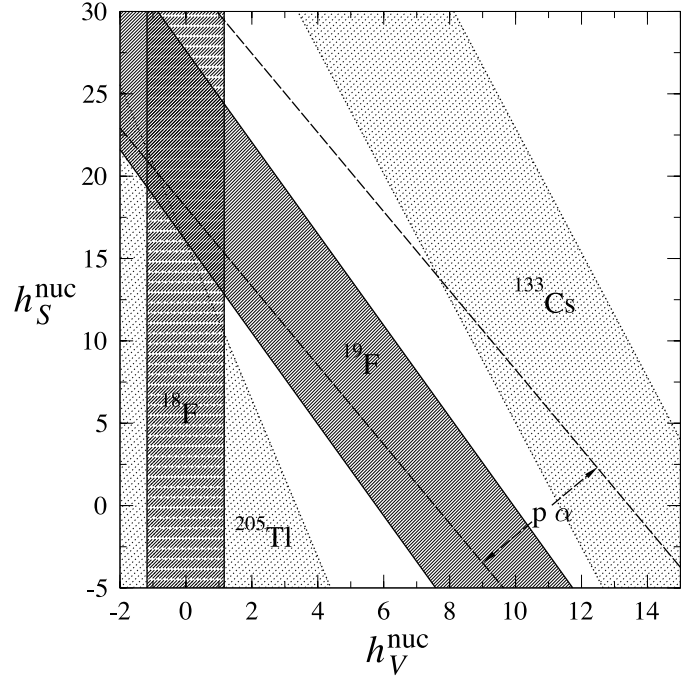}}
   \caption{Constraints on effective DDH weak meson-nucleon coupling
    constants deduced from parity violating observables in light nuclei and
    anapole moments of heavy atoms (see [5]). The combinations of the weak
    meson-nucleon coupling constants are:
    $h^{nuc}_{V} = h^{1}_{\pi} - 0.12h^{1}_{\rho} - 0.18h^{1}_{\omega}$ and
    $h^{nuc}_{S} = -(h^{0}_{\rho} + 0.7h^{0}_{\omega})$}
   \label{zfig1}
   \end{figure}
   More straightforward in terms of interpretation are the
    measurements of the proton-proton parity violating analyzing powers or
    asymmetries (the normalized differences in scattering yields for positive
    and negative helicity incident beams) for which there exist three
    precision experiments (from the University of Bonn at 13.6 MeV [6],
    from PSI at 45 MeV [7], and from TRIUMF at 221 MeV [8]). These
    experiments required unprecedented precision in controlling systematic
    errors and in particular those due to helicity correlated false
    asymmetries. The parity violating analyzing powers are compared
    to more recent theoretical calculations in the framework of weak
    meson-nucleon couplings in Fig. 2. (see [8])
 \begin{figure}
   \begin{center}
   \includegraphics*[width=0.9\linewidth]{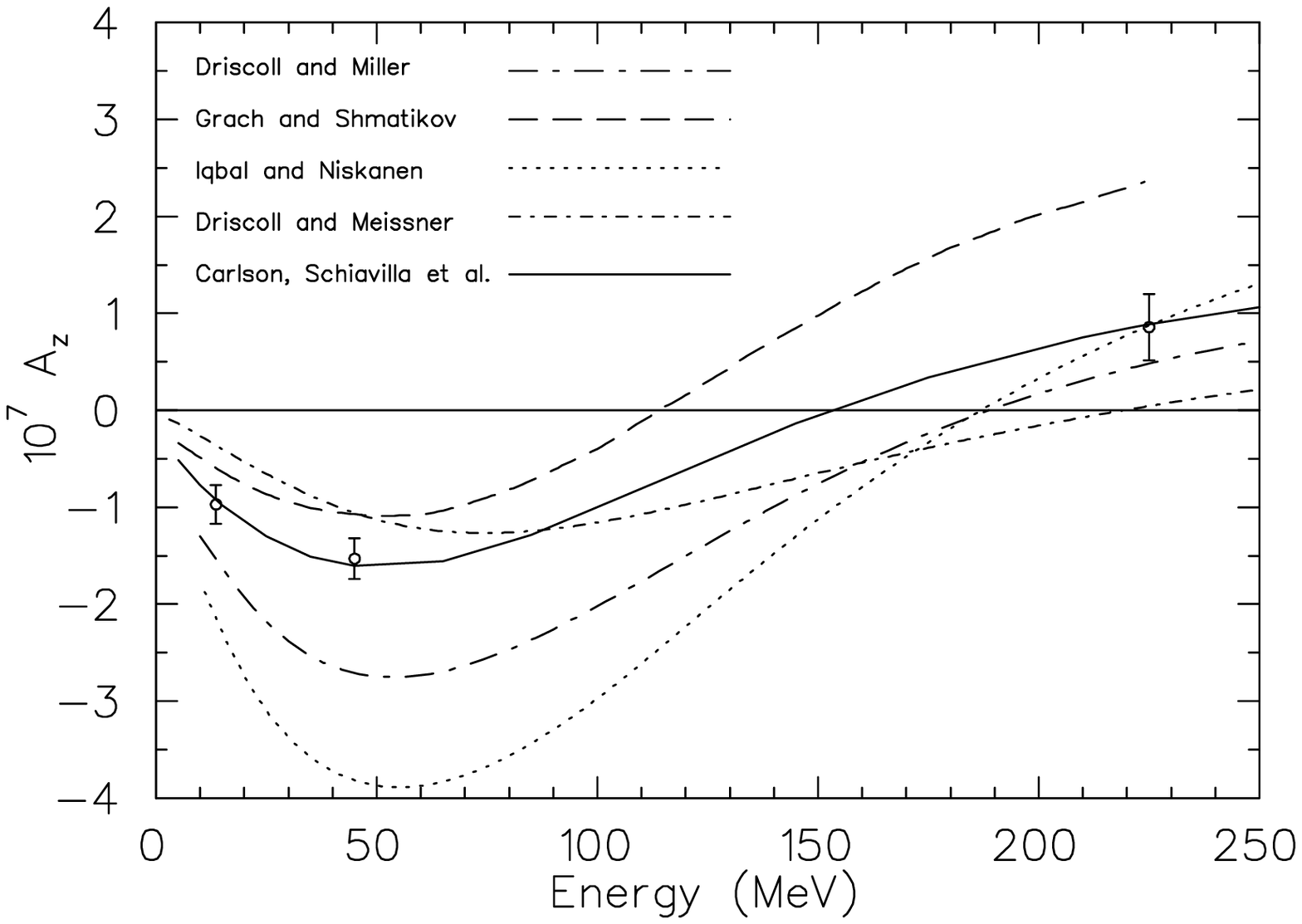}
   \end{center}
   \caption{The three most precise proton-proton parity violation
    measurements (at 13.6, 45, and 221 MeV) and theoretical predictions.
    The solid curve shows the calculation by Carlson et al.[9] in which
    the weak meson nucleon coupling constants were fitted to the data.}
   \label{zfig2}
   \end{figure}
    Carlson et al. [9] deduced the constraints on the two combinations of
    weak meson-nucleon coupling constants:
    $h^{pp}_{\omega} = h^{0}_{\omega} + h^{1}_{\omega}$ and
     $h^{pp}_{\rho} = (h^{0}_{\rho} + h^{1}_{\rho} + h^{2}_{\rho}/\sqrt{6})$
    from the proton-proton analyzing power data. Clearly even more precise
    measurements of the proton-proton analyzing powers are required in order
    to better constrain the weak meson-nucleon coupling constants (see Fig. 3).
   \begin{figure}
 \centerline{\includegraphics*[width=0.8\linewidth]{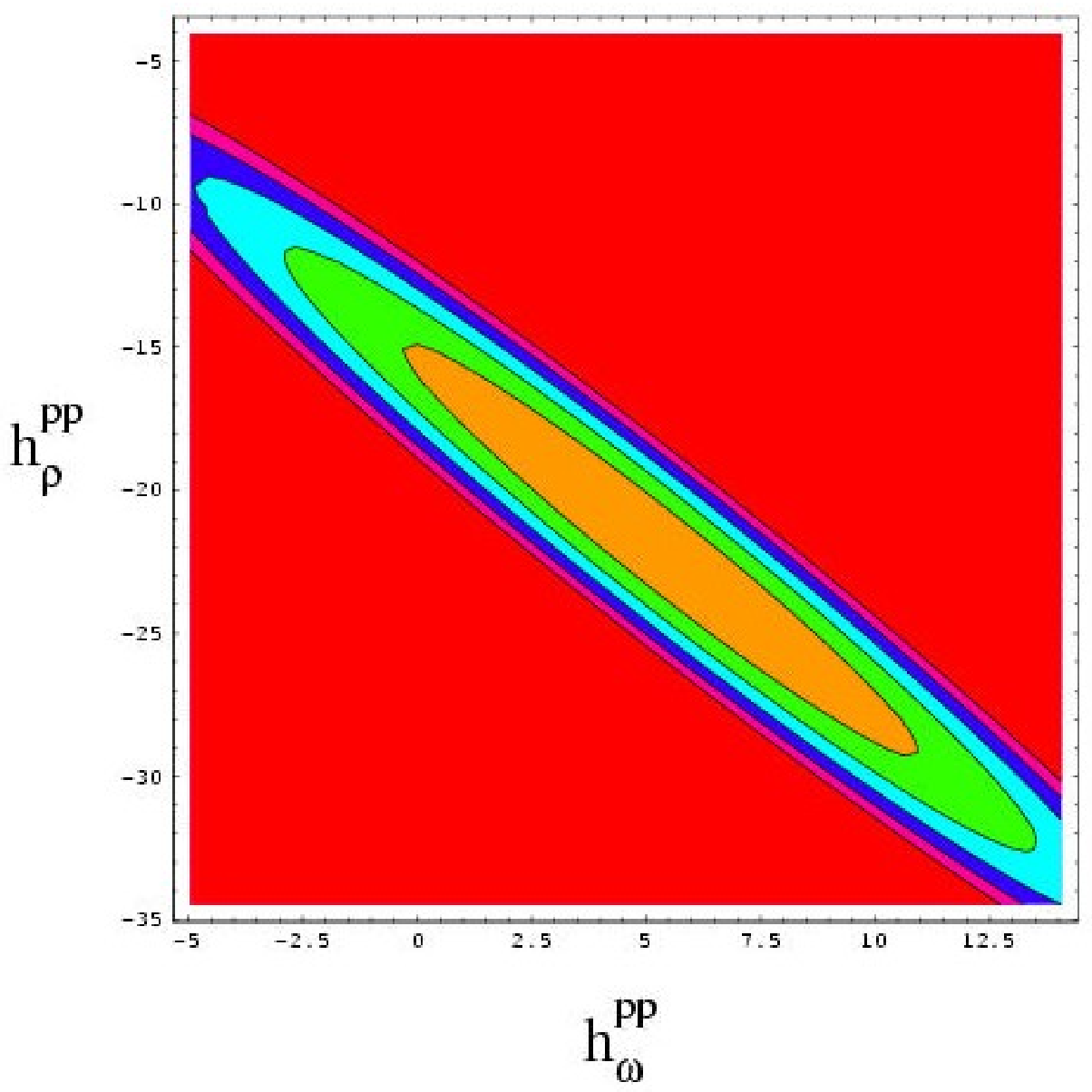}} 
 \caption{Constraints on the weak meson-nucleon coupling constants,
    combinations $h^{pp}_{\omega}$ and $h^{pp}_{\rho}$ as deduced from the
    proton-proton parity violating analyzing powers. [9] The figure shows
    contours of constant total $\chi^{2} = 1, 2, 3, 4, 5$. The axes scales
    are in units of $10^{-7}$.}
   \label{zfig3}
   \end{figure}

   The extensive development of chiral perturbation theory and N-N effective
   field theory and substantial progress in performing lattice QCD
   calculations have opened up new approaches of studying the hadronic weak
   interaction. For a recent review see Ref. 10. In principle it is now
   possible to make QCD based predictions for weak hadronic interaction
   phenomena. On the experimental side there are several promising
   precision parity violation measurements in progress. Foremost to be
   mentioned is the measurement of the parity violating asymmetry in the
   capture of polarized neutrons on hydrogen (the NPDGAMMA experiment).
   Others are the parity violating neutron spin rotation in helium and in
   hydrogen. (see Ref. 10) 
   
\section{Parity Violating Electron Scattering}

   The situation with regard to parity violating analyzing power
   measurements in polarized electron scattering is quite different.
   Originally, in the low energy regime, these measurements were made to
   determine the weak mixing angle or $\sin^{2}\theta_{W}$. Soon thereafter
   these measurements were overtaken by measurements at the $Z^{0}$ pole
   at LEP of CERN and SLC of SLAC. Given the current knowledge of the
   electroweak interaction and the great precision with which electroweak
   radiative corrections can be calculated, the current
   round of parity violating electron scattering experiments has as one
   of its objectives to probe the structure of the nucleon and in particular
   the contributions of the strange quarks (the quark-antiquark pairs of the
   sea) to the weak charge and magnetization distributions of the nucleons.
   With the couplings of both photons and Z bosons to point-like quarks 
   well defined (see Table 2), it is possible to separate the contributions
   of the various flavors. The electromagnetic and weak charge and magnetic
   form factors of the proton can be written:
   \begin{eqnarray*}
   G^{\gamma,Z~p}_{E,M} = 2q^{u}G^{u}_{E,M} + q^{d}G^{d}_{E,M} + q^{d}G^{s}_{E,M}
    \end{eqnarray*}
   Heavier flavor contributions are neglected. Assuming charge symmetry in
   the quark distributions, one can write an analogous expression for the
   neutron:
    \begin{eqnarray*}
    G^{\gamma,Z~n}_{E,M} = 2q^{d}G^{d}_{E,M} + q^{u}G^{u}_{E,M} + q^{d}G^{s}_{E,M}
    \end{eqnarray*}
   Clearly, in addition to the electromagnetic form factors for
   the proton and the neutron one needs one further relation, which is
   provided by the parity violating analyzing power $A_{z}$:
    \begin{eqnarray*}
   A_{z} &=& (1/P)[N^{+} - N^{-}]/[N^{+} + N^{-}] \\ 
    \mbox{} &=&  -[G_{F}Q^{2}/(4\sqrt{2}\pi\alpha)]\times[\epsilon G^{\gamma}_{E}G^{Z}_{E}
 + \tau G^{\gamma}_{M}G^{Z}_{M} - (1 - 4 \sin^{2}\theta_{W})\epsilon 'G^{\gamma}_
{M}G^{e}_{A}]/D
\end{eqnarray*}
where $N^{+}$ and $N^{-}$ are the normalized scattering yields and
\begin{eqnarray*}
        \tau     &=& Q^{2}/(4M^{2}_{p})\, ,\\
   \epsilon   &=& 1/(1 + 2(1 + \tau )\tan^{2}(\theta/2))\, ,\\
                 D&=& \epsilon (G^{\gamma}_{E})^{2} + \tau (G^{\gamma}_{M})^{2}\, ,\\
   \epsilon ' &=& \sqrt{\tau (1 + \tau )(1 - \epsilon ^{2})}\, ,
\end{eqnarray*}
   and $Q^{2}$ is the squared four-momentum transfer, $G_{F}$ is the Fermi
   coupling constant and $\alpha$ is the fine structure constant. The three new
   form factors $G^{Z}_{E}$, $G^{Z}_{M}$, and $G^{e}_{A}$ can be determined by
   measuring parity violating electron elastic scattering from the proton at
   forward and backward angles, and quasi-elastic scattering from the deuteron
   at backward angles.
   
 \begin{table}[h]
\tbl{Electroweak charge phenomenology. Note the accidental suppression
 of the weak charge of the proton. The two charges of the proton and the
 neutron are to a large extent interchanged.}
 {\begin{tabular}{@{}lcr@{}} \toprule
 & Electromagnetic Charge & Weak Charge \\
\colrule
 $q^{u}$  &  $+2/3$  &  $ 1 - (8/3) \sin^{2}\theta_{W} \approx  1/3$\\
 $q^{d}$  &  $-1/3$  &  $-1 + (4/3) \sin^{2}\theta_{W} \approx -2/3$\\
 $Q^{p} = 2q^{u} + 1q^{d}$  &  $+1$  &  $1 - 4 \sin^{2}\theta_{W} = 0.0716$\\
 $Q^{n} = 1q^{u} + 2q^{d}$  &  $0$   &  $-1$\\ \botrule
 \end{tabular}}
 \end{table}

\begin{figure}[t!]
\centerline{\includegraphics*[width=0.8\linewidth]{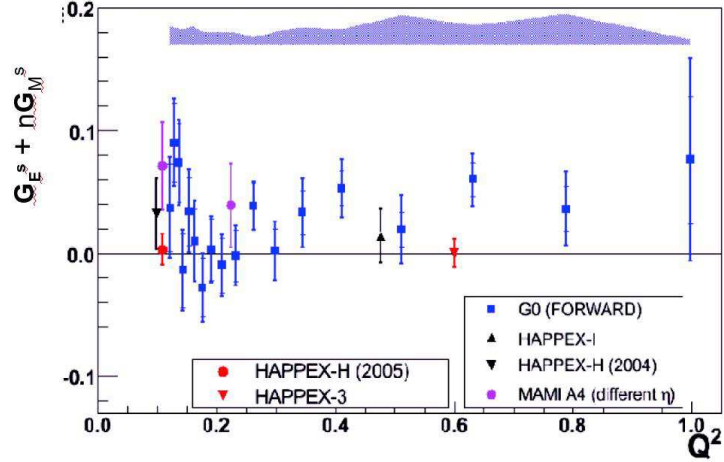}}
 \caption{The combination $G^{s}_{E} + \eta G^{s}_{M}$ as deduced from the G0,
 HAPPEX, and PVA4 experiments. The blue-gray band indicates the dominant
 systematic uncertainty in the G0 results.}
\label{zfig4}
\end{figure}    

   Presently there exist published data on a combination
   $G^{s}_{E} + \eta G^{S}_{M}$, with $\eta $ a coefficient dependent on
   the four-momentum transfer, from the SAMPLE experiment at MIT-Bates
   [11], the HAPPEX [12] and G0 [13] experiments at JLab, and the PVA4
   experiment at MAMI 14 (see Fig. 4). There is consistency between the
   various experimental results. The data indicate tiny non-zero values
   for the combination $G^{s}_{E} + \eta G^{S}_{M}$. The backward angle
   data on hydrogen and deuterium of the G0 experiment are being analyzed,
   while further measurements are on the schedules of JLab (HAPPEX) and
   of MAMI (PVA4). Extrapolating the forward angle results towards zero
   four-momentum transfer and zero scattering angle, has provided a factor of five improvement in
   the knowledge of the neutral weak couplings to the `up' and `down'
   quarks [15] compared to the entries in the PDG Handbook of 2006.

\begin{figure}
\centerline{\includegraphics*[width=0.7\linewidth]{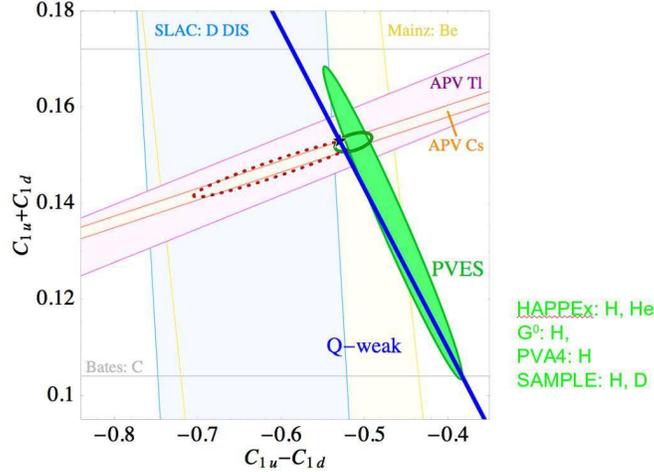}}
 \caption{The neutral weak effective couplings to the quarks. The dotted
 contour displays the previous experimental limits (95\%) as reported
 by the PDG in 2006 together with the prediction of the Standard Model
 (black star). The filled ellipse gives the new constraint provided by
 the recent parity violation electron scattering measurements on hydrogen,
 deuterium, and helium (at 1 standard deviation). The solid contour (95\%
  CL) presents the constraint based on all experimental results. All other
  experimental limits shown are displayed at 1 standard deviation. The
heavy line represents the anticipated constraint imposed by the Qweak
  experiment at JLab assuming that the experimental result agrees with the
  Standard Model.}
\end{figure}

\section{Tests of the Standard Model}

    In going to lower and lower four-momentum transfer and small scattering angle in parity
   violating electron elastic scattering from the proton, the contributions
   due to the finite size of the proton become smaller and smaller and one
   is able to measure then the weak charge of the proton, which constitutes
   the sum of the weak charges of the two `up' quarks and the `down' quark.
   However, the analyzing power becomes zero at zero momentum transfer and
   therefore optimum values of four-momentum transfer and incident energy
   need to be sought. A high precision measurement of the parity violating
   analyzing power determines the value of $\sin^{2}\theta_{W}$ and
   consequently the variation of $\sin^{2}\theta_{W}$ with four-momentum
   transfer or the `running' of $\sin^{2}\theta_{W}$. The Standard Model
   makes a definitive prediction of the `running' of $\sin^{2}\theta_{W}$
   taking into account electroweak radiative corrections once the value of
   $\sin^{2}\theta_{W}$ at the $Z^{0}$ pole has been reproduced. As with
   the QED and QCD couplings, $\alpha(\mu^{2})$ and $\alpha_{s}(\mu^{2})$
   (which exhibit screening and anti-screening, respectively), in going to
   higher and higher four-momentum transfer, $\sin^{2}\theta_{W}$ is an
   effective parameter also varying with $\mu^{2} \approx Q^{2}$. In this
   case the behaviour with $Q^{2}$ is more subtle since $\sin^{2}\theta_{W}$
   is a function of the electroweak couplings $g_{Vl}$ and $g_{Al}$:
   $(g_{Vl}/g_{Al}) = 1 - 4\sin^{2}\theta_{W}$. Any deviation of
   $\sin^{2}\theta_{W}$ from its Standard Model value points to new physics,
   which needs to be incorporated through a set of new diagrams. In going
   to very high energies it is plausible that the strong, electromagnetic,
   and weak interactions become the same. In fact extrapolations to very
   high energies of the three coupling constants show a `near-miss' towards
   having a common intersection (see Fig. 6). Extensions of the Standard
   Model are therefore in order.
  
\begin{figure}
\centerline{\includegraphics*[width=0.8\linewidth]{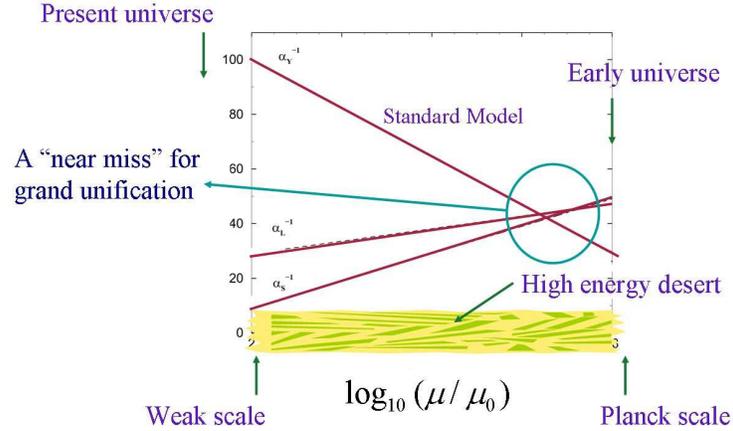}}
 \caption{Extrapolation of the reciprocals of the three coupling constants
 (electromagnetic, weak, and strong interaction) to the GUT scale}
\end{figure}
 
   Measurements at the $Z^{0}$ pole have established
   the value of $\sin^{2}\theta_{W}$ with great precision although it must
   be remarked that
   the leptonic and semi-leptonic values of $\sin^{2}\theta_{W}$ differ by
   $3\sigma$. The Standard Model `running' of $\sin^{2}\theta_{W}$ has been
   calculated by Erler, Kurylov, and Ramsey-Musolf [16] in the modified
   minimal subtraction scheme (see Fig. 7). The theoretical uncertainties
   in the `running' of $\sin^{2}\theta_{W}$ are represented by the width of
   the curve. Hence the interpretability is currently limited by the
   normalization of the curve at the $Z^{0}$ pole, which is arguably as
   small as $\pm0.00016$. Note the shift of $+0.007$ at low $Q^{2}$ with
   respect to   the $Z^{0}$ pole best fit value of $0.23113 \pm 0.00015$.
   There have been reported several low energy measurements of the value
   of $\sin^{2}\theta_{W}$. The first one is from an atomic parity
   violation measurement in Cesium [17], which agrees with the
   Standard Model prediction within $1\sigma$ after many refinements
   detailing the atomic structure of Cesium were introduced. The
   second one is from a measurement of parity violating M{\o}ller
   scattering [18], which also agrees with the Standard Model prediction
   within approximately $1\sigma$. This is at present the better
   measurement in constraining extensions of the Standard Model. The third
   one is from a measurement of neutrino and antineutrino scattering from
   iron [19] with a roughly $3\sigma$ deviation from the Standard Model
   prediction. For this result there remain various uncertainties in the
   theoretical corrections that need to be applied (among other two
   distinct effects of charge symmetry breaking in the quark distributions
   of the nucleons [20]). It is quite apparent that much higher precision
   experiments are needed in order to search for possible extensions of
   the Standard Model. One of these is a precision measurement of the weak
   charge of the proton,
   $Q^{p}_{W} = 1 - 4\sin^{2}\theta_{W}$, currently being prepared for
   execution starting in 2009 in Hall C of Jefferson Laboratory [21]. The
   extraction of the value of $\sin^{2}\theta_{W}$ is free of many-body
   theoretical uncertainties and has the virtue of being able to reach much
   higher precision (note that at a $Q^{2}$ value of $0.03~\textrm{(GeV/c)}^{2}$,
   $1 - 4\sin^{2}\theta_{W}$ equals 0.07 giving a large boost in the precision
   that may be obtained). The dominant hydronic effects that must be
   accounted for in extracting $Q^{p}_{W}$ from the measured analyzing power
   are contained in form factor contributions which are sufficiently
   constrained from the recent programs of parity violating electron
   scattering (at MIT-Bates, JLab, and MAMI) without reliance on theoretical
   nucleon structure calculations, with the exception possibly of very small
   two-photon exchange contributions). The Standard Model evolution of
   $\sin^{2}\theta_{W}$ corresponds to a 10 standard deviation effect in the
   planned Qweak experiment at JLab. The Qweak experiment, the first ever
   precision measurement of the weak charge of the proton and more precise
   than the existing low energy measurements, is crucial in testing the
   Standard Model.

\begin{figure}
\centerline{\includegraphics*[width=0.7\linewidth]{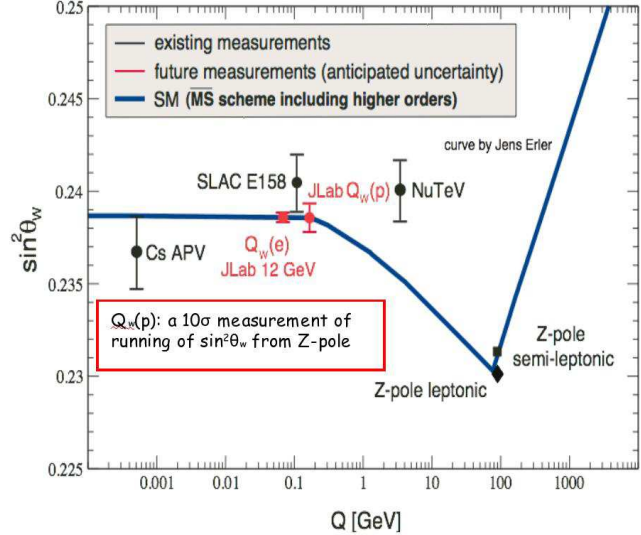}}
 \caption{Calculated `running' of the weak mixing angle in the Standard Model,
 as defined in the minimal subtraction scheme. [16] The black points with 
 $1\sigma$ error bars show the existing experimental values, while the red
 points with error bars refer to the 4\% $Q^{p}_{W}$ measurement in
 preparation and a 2.5\% $1$~GeV M{\o}ller measurement under consideration.}
\end{figure}

\begin{figure}[t!]
\centerline{\includegraphics[width= \linewidth]{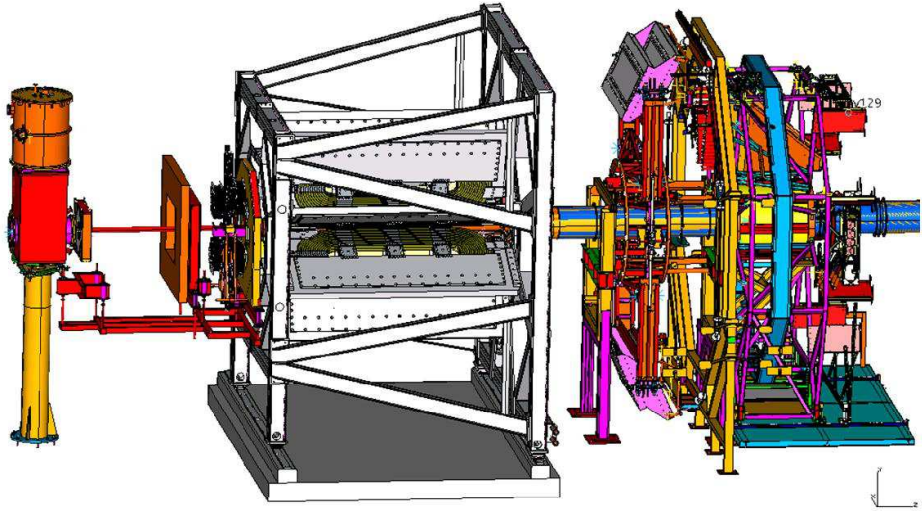}}
\caption{Layout of the Qweak experiment. The beam is incident from the left 
 and scattered electrons exit the target and pass through the first
 collimator, the region-1 GEM detectors, the two stage second precision
 collimator which surrounds the region-2 drift chambers, the toroidal magnet,
 the shielding wall, the region-3 drift chambers , the trigger scintillators,
 and finally the ersatz quartz \v{C}erenkov detectors. The tracking system
 chambers and the trigger scintillators, mounted on rotatable wheels, will 
 be retracted outwards during high current data taking to measure $A_{z}$.
 The luminosity monitors, which will be used to monitor target density
 fluctuations and to provide sensitive null tests, are located downstream
 of the main apparatus very close to the through going beam. Further luminosity monitors are placed just downstream of the LH$_2$ target. }
\end{figure}

   In the Qweak experiment the weak charge of the proton,
   $Q^{p}_{W} = 1 - 4\sin^{2}\theta_{W}$, will be deduced from the parity
   violating analyzing power, defined as:
 \begin{eqnarray*}
   A_{z} = (1/P)[N^{+} - N^{-}]/[N^{+} + N^{-}]
   \end{eqnarray*}
   where P is the polarization of the longitudinally polarized electron beam
   and $N^{+}$ and $N^{-}$ are the normalized scattering yields.
   It was shown in [22] that for forward angle scattering, where
   $\theta \rightarrow 0$, the analyzing power can be written:
   \begin{eqnarray*}
   A_{z} = (-G_{F}/(4\sqrt{2}\pi\alpha))[Q^{2}Q^{p}_{W} + Q^{4}B(Q^{2})]
   \end{eqnarray*}
   Here $G_{F}$ denotes the Fermi coupling constant and $\alpha$ is the fine
   structure constant. One should note the dependence on $P$, which requires
   precision polarimetry, and the dependence on the average value of $Q^{2}$
   over the finite acceptance of the magnetic spectrometer based detector
   system for scattered electrons, which requires the averaged value to be
   determined through specific ancillary control measurements. The leading
   term in the equation is the weak charge of the proton,
   $Q^{p}_{W} = 1 - 4\sin^{2}\theta_{W}$. The quantity $B(Q^{2})$ represents
   the finite size nucleon structure and contains the proton and neutron
   electromagnetic and weak form factors. The value of $B(Q^{2})$ can be
   determined experimentally by extrapolation from the ongoing program of
   forward angle parity violating electron scattering experiments at higher
   values of $Q^{2}$, discussed above, or by specific control measurements.
   The incident energy and the four-momentum transfer value (mean scattering
   angle) followed from careful considerations of the figure of merit. The
   optimum values are an incident energy of 1.165~GeV and a four-momentum
   transfer of $0.03~\textrm{(GeV/c)}^{2}$. One can then write for the longitudinal
   analyzing power:
   \begin{eqnarray*}
   A_{z}(0.03~\textrm{(GeV/c)}^{2})  &=& A(Q^{p}_{W}) + A(Had_{V}) + A(Had_{A}) \\
                                              &=& -0.19 ppm - 0.09 ppm - 0.01 ppm
   \end{eqnarray*}
   where the hadronic structure contributions are separated in vector and
   axial vector components. Clearly, the analyzing power is very small 
   $(-0.3 ppm)$ and one must arrive at an overall uncertainty of 2\%
   to meet the precision objective of 0.3\% in $\sin^{2}\theta_{W}$.
   Consequently, high statistics data are a prerequisite requiring high
   luminosity and high beam polarization, and an integrating low-noise 
   detector system of large acceptance. As indicated above, the longitudinal
   beam polarization must be precisely known as well as the hadronic
   structure contribution $B(Q^{2})$ to be subtracted from the measured
   analyzing power $(A(Had_{V}) + A(Had_{A}))$ The Qweak experiment requires
   2200 hours of data taking of the longitudinal analyzing power $A_{z}$ in
   elastic electron proton scattering at a momentum transfer of
   $0.03~\textrm{(GeV/c)}^{2}$ with 180 $\mu$A of 85\% polarized beam incident on
   a 0.35~m long LH$_{2}$ target to determine $\sin^{2}\theta_{W}$ at the
   0.03\% level at low $Q^{2}$. A layout of the Qweak experiment is given
   in Fig. 8.

   The Qweak experiment is complementary to a parity violating electron-electron 
    scattering experiment, under consideration to be performed at
   11~GeV with an upgraded CEBAF at JLab, with an envisaged precision in
   $\sin^{2}\theta_{W}$ equal to or better than that from any individual
   measurement at the $Z^{0}$ pole. This implies a considerable reduction
   in the error of the SLAC M{\o}ller experiment. Needless to state:
   the electroweak radiative corrections to a pure leptonic measurement are
   more contained. In the search for physics beyond the Standard Model,
   precision measurements of the weak charge of the proton and of the weak
   charge of the electron are rather complementary.
 
 \section{Acknowledgement}
 The author would like to thank W. Desmond Ramsay for critical reading of the manuscript.

\end{document}